\pdfoutput=1
\documentclass{JINST}
\usepackage{textcomp} 
\DeclareGraphicsExtensions{.png,.pdf,.jpg} 
\usepackage{ifpdf}
\usepackage{lineno}

\title{Trigger and readout electronics for the STEREO experiment.}

\author{O.~Bourrion$^a$\thanks{Corresponding author},
J.\,L.~Bouly$^a$,
J.~Bouvier$^a$,
G.~Bosson$^a$,
V.~Helaine$^a$,
J.~Lamblin$^a$,
C.~Li$^a$,
F.~Montanet$^a$,
J.\,S.~Real$^a$,
T.~Salagnac$^a$,
N.~Ponchant$^a$,
A.~Stutz$^a$,
D.~Tourres$^a$,
C.~Vescovi$^a$,
S.~Zsoldos$^a$.\\
\llap{$^a$}LPSC, Universit\'e Grenoble-Alpes, CNRS/IN2P3 \\
53, rue des Martyrs, Grenoble, France}

\abstract{The STEREO experiment will search for a sterile neutrino by measuring the anti-neutrino energy spectrum as a function of the distance from the source, the ILL nuclear reactor. 
A dedicated electronic system, hosted in a single microTCA crate, was designed for this experiment. It performs triggering in two stages with various selectable conditions, processing and readout via UDP/IPBUS of 68 photomultiplier signals continuously digitized at 250\,MSPS. Additionally, for detector performance monitoring, the electronics allow on-line calibration by driving LED synchronously with the data acquisition.
This paper describes the electronics requirements, architecture and the performances achieved.}


\begin{document}
\setpagewiselinenumbers

\section{Introduction}
\label{intro}
The STEREO experiment \cite{stereoRef1,stereopost} searches for a sterile neutrino by measuring the anti-neutrino energy spectrum as a function of the distance from the source, the core of the Institut Laue-Langevin (ILL) research nuclear reactor.

This measurement will be done using the interaction of the anti-neutrino in a liquid scintillator (LS) via the inverse beta decay process (IBD).
The anti-neutrino signature relies on the coincidence of a positron interaction and the delayed neutron capture within a few 10\,\textmu s (exponential decay).

The target of the detector is filled with a gadolinium-loaded LS. 
The volume is optically segmented in six optically-separated cells, each containing four photomultipliers (PMT).
The target volume is surrounded by an outer crown called ``Gamma-catcher'' and filled with LS without gadolinium.
The outer crown recovers part of the escaping gammas to improve the detection efficiency and the energy resolution. The Gamma-catcher is viewed by 24 PMT.

For each anti-neutrino interaction occurring in a detector cell, each PMT can collect up to 1500 photo-electrons.
\begin{figure}
\begin{center}
\includegraphics[angle=-90,width=0.7\textwidth]{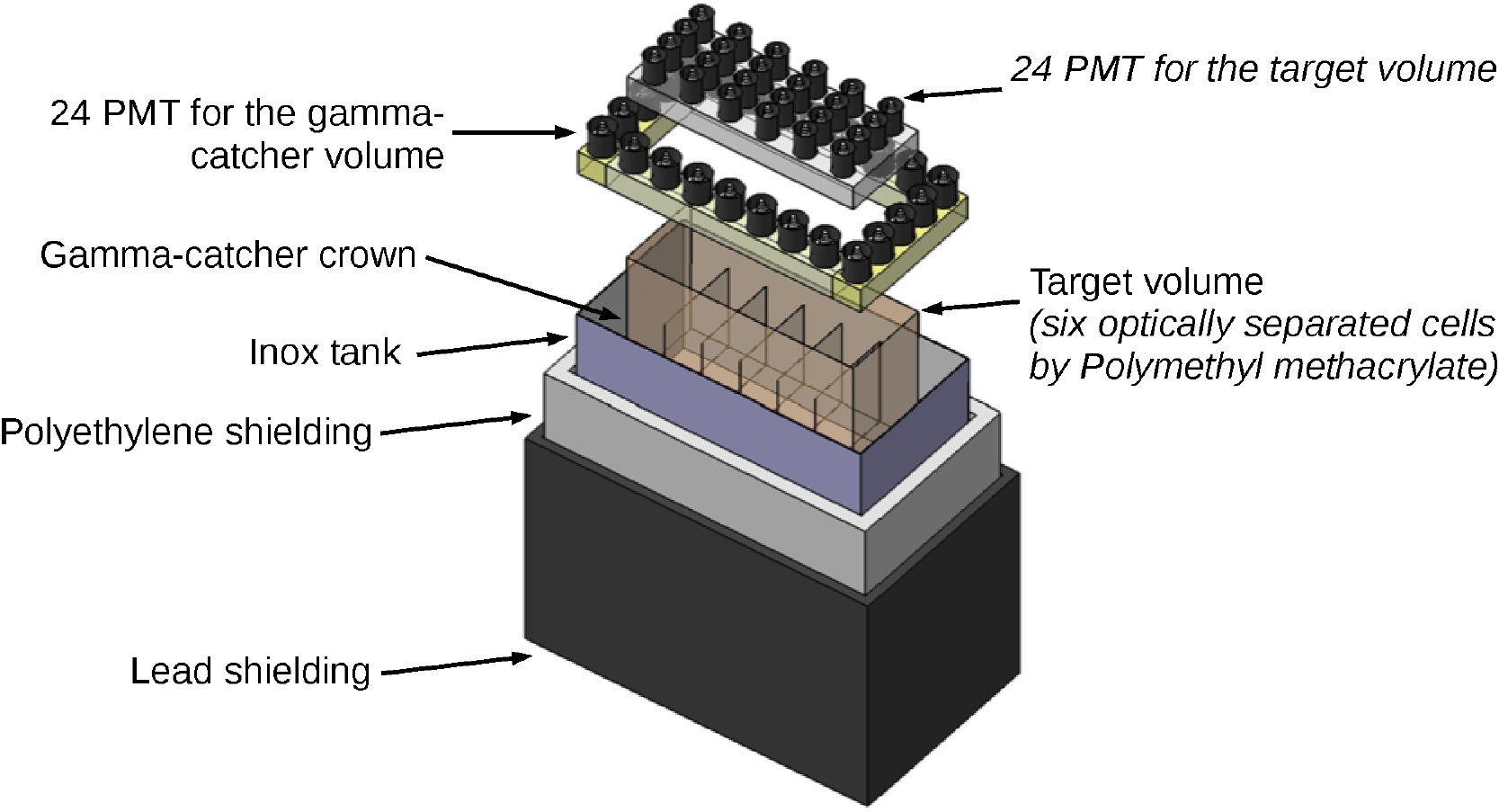}
\caption{Exploded view of the STEREO detector (dimensions  $\rm 3\,m \times 1\,m \times 1.5\,m$). The Cerenkov detector used to reject cosmic background events is not shown.}
\label{stereoExploded}
\end{center}
\end{figure}
In addition, in order to reject cosmic background events, a Cerenkov detector containing 20 PMT is located above the whole detector.

For the three parts of the STEREO detector, the total duration of the expected light signals is between 100 and 200\,ns and the fast rise part of the signal is lasting about 20\,ns.
The event rate for the whole detector could be as high as 1\,kHz.

The STEREO detector is regularly calibrated with a LED system.
The light produced by different and independent LED boxes, each containing 6 LED, is injected at different points in the detector thanks to optical fibers.

\section{Electronics requirements overview}
\label{overview}
\begin{figure}
\begin{center}
\includegraphics[angle=-90,width=0.65\textwidth]{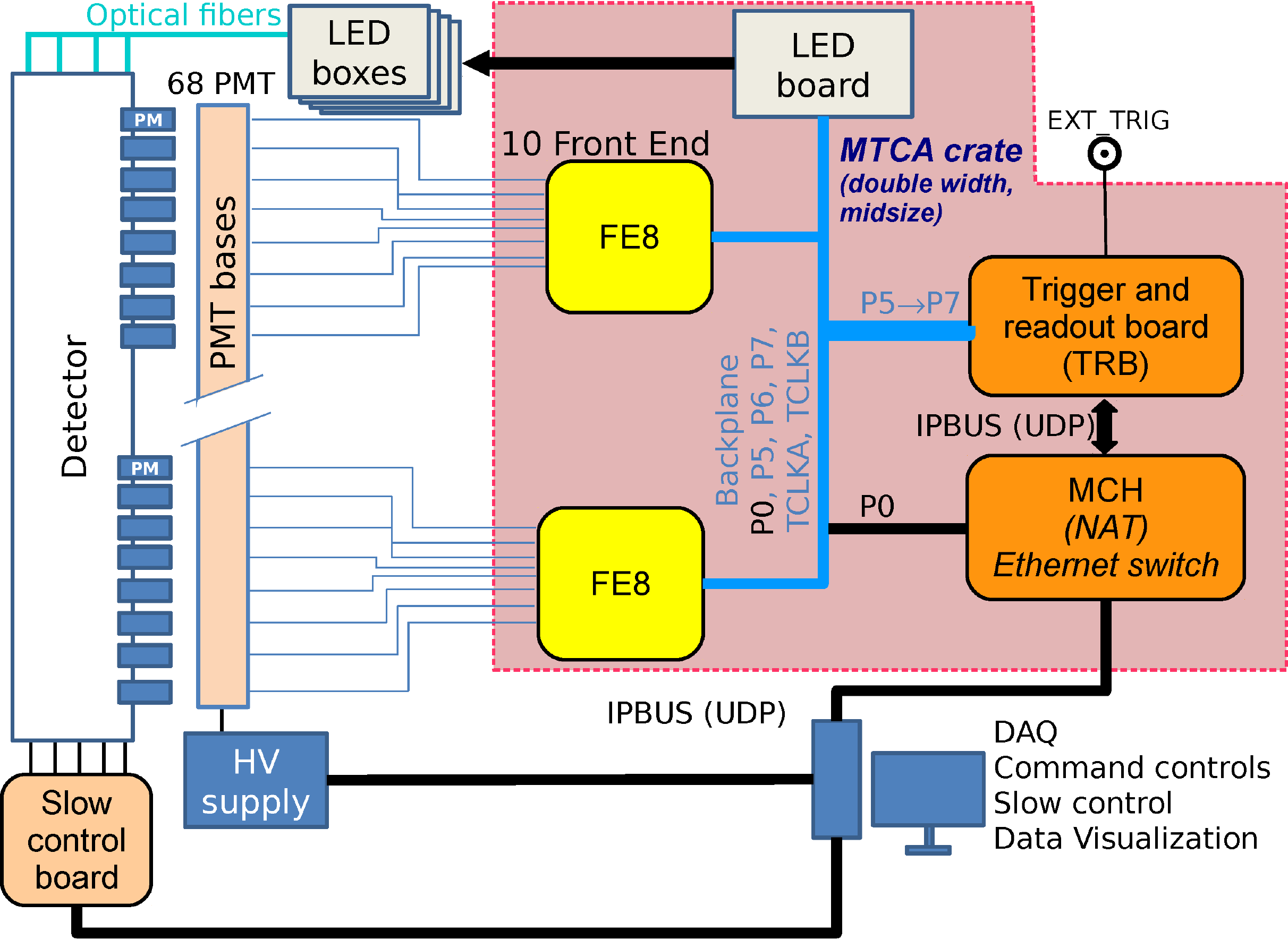}
\caption{STEREO experiment DAQ electronics overview. The microTCA crate is equipped with ten FE8 boards, one TRB board and one LED board. Readout and slow control are done by IPBUS.}
\label{electronicsOverview}
\end{center}
\end{figure}

A dedicated electronic system, hosted in a single microTCA (MTCA) crate, was designed for the STEREO experiment, see figure~\ref{electronicsOverview}. 
It serves several purposes: triggering, processing, readout and on-line calibration. 
For that purpose, the MTCA crate is equipped with ten 8-channels front-end electronic boards (FE8), one trigger and readout mezzanine board (TRB) mounted on the MicroTCA Carrier Hub (MCH) and one LED board to drive the LED boxes.

\begin{figure}
\begin{center}
\includegraphics[angle=0,width=0.55\textwidth]{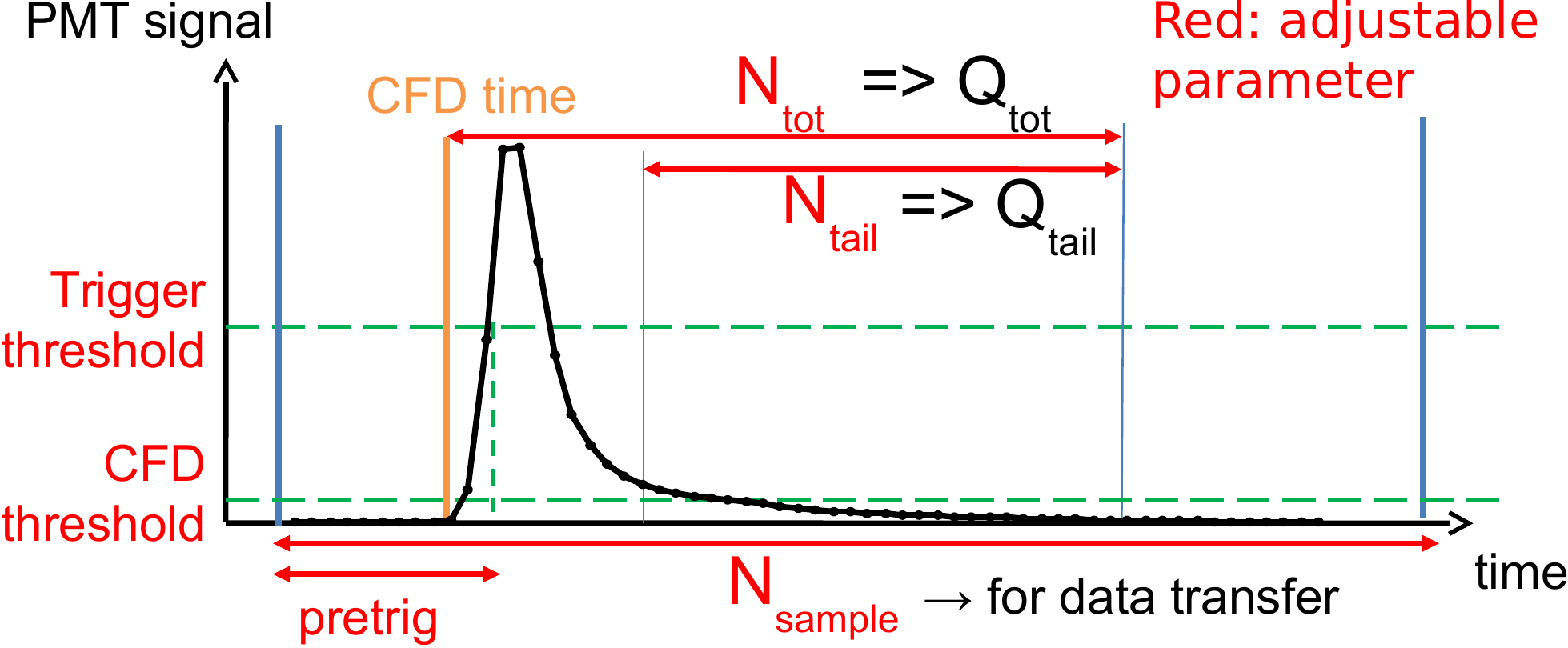}
\caption{Illustration of the first level trigger generation and of the pulse shape discrimination. A real signal pulse is shown, each 4\,ns sample is represented by a dot.}
\label{PSD_1}
\end{center}
\end{figure}
The FE8 are in charge of continuously digitizing the 68 PMT signals at 250\,MSPS for performing two main tasks that are illustrated in figure~\ref{PSD_1}. 
At first, a FE8 can generate a candidate trigger (or a veto) if any of its channel is above the trigger threshold.
This candidate trigger is used by the TRB to build, depending on the selected trigger conditions, a first level accepted trigger (T1a) that is returned to the FE8 in 124\,ns.
In the second step, when the confirmed trigger is received, each FE8 performs signal processing.

For that, the beginning of the signal is to be found with a Constant Fraction Discriminator (CFD) having its own threshold set above noise, 
this search is done in a window having a length of $\rm N_{sample}$ samples (see figure~\ref{PSD_1}).
The total charge $\rm Q_{tot}$ and the tail charge  $\rm Q_{tail}$, useful for the pulse shape discrimination (PSD) method, are respectively the results of the Riemann integration over $\rm N_{tot}$ samples (typical total pulse duration) 
and $\rm N_{tail}$ samples (typical pulse tail duration).

The TRB is used to perform the second level trigger (T2), to collect and aggregate the processed data provided by the FE8.
The  TRB is also used to drive the LED board during calibrations.
The TRB is installed on a commercial MicroTCA Carrier Hub (MCH) from NAT\textregistered \  \cite{NAT} and provides the system clock (250\,MHz) to all AMC slots, i.e. the FE8 and the LED boards. 
The amount of light generated by the LED is set by the LED board which is configured by slow control.

The communication between the electronic boards is done via custom serial protocols. 
The slow control and the data acquisition is done by a modified version of the IPBUS \cite{IPBUS}, that allows the Dynamic Host Configuration Protocol (DHCP).
For the reader comfort, it may be added that the standard User Datagram Protocol (UDP) protocol\cite{UDP}, which is a very simple Ethernet protocol, is unreliable by definition.
Consequently, the IPBUS was designed as a UDP enhancement aimed at adding a reliability mechanism.
As the IPBUS protocol is simple, it can be implemented directly in FPGA, without the need of having a Central Processing Unit.

\section{Front-end board}
\label{FEB}
\begin{figure}
\begin{center}
\includegraphics[angle=-0,width=0.5\textwidth]{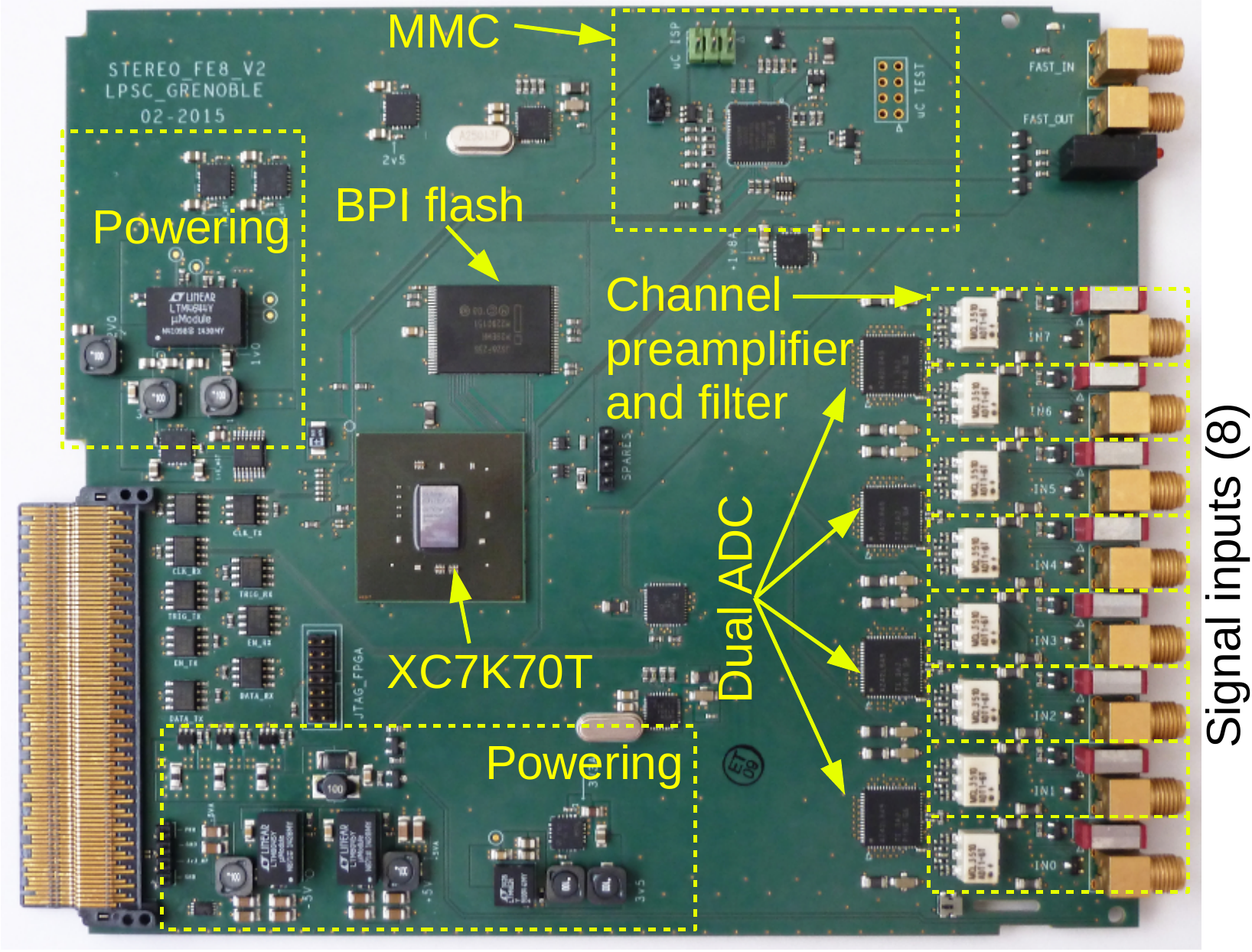}
\caption{Picture of the front-end electronic board (FE8).}
\label{FE8pic}
\end{center}
\end{figure}

The front-end electronic board, shown in fig.~\ref{FE8pic}, features 8 analog inputs that are pre-amplified with two selectable gains ($\times 1$ and $\times 20$).
The second gain was implemented to allow the single photo-electron separation, hence permitting PMT calibration with LED.
The amplifiers outputs are AC coupled ($\rm f_c >30\,kHz$) and equipped with 5\textsuperscript{th} order anti-aliasing filters ($\rm f_{-3dB}=85\,MHz$).
Finally, the 8 filtered signals are sampled by four dual 14-bit ADC (ADS42LB49) operated at 250\,MSPS.
By comparing the charge of simulated pulses before and after the sampling, we ensured that the selected signal sampling rate is sufficient to permit an accurate charge calculation and to allow a pulse-shape discrimination (PSD) which fulfills the STEREO requirements.

ADC are controlled and read out by a single FPGA (XC7K70T-2TFBG676).
This FPGA is also in charge of providing the IPBus connectivity with the DAQ and the trigger and readout communication links with TRB.
It must be noted that the FPGA is configured via the Boot Parallel Interface mode (BPI) and that the firmware stored in the flash memory can be updated via IPBus.
As required by the standard \cite{MTCA}, the board features a Module Management Controller (MMC).
The MMC implemented is a modified version of the one distributed by CERN \cite{MMC}.

\begin{figure}
\begin{center}
\includegraphics[angle=-0,width=0.8\textwidth]{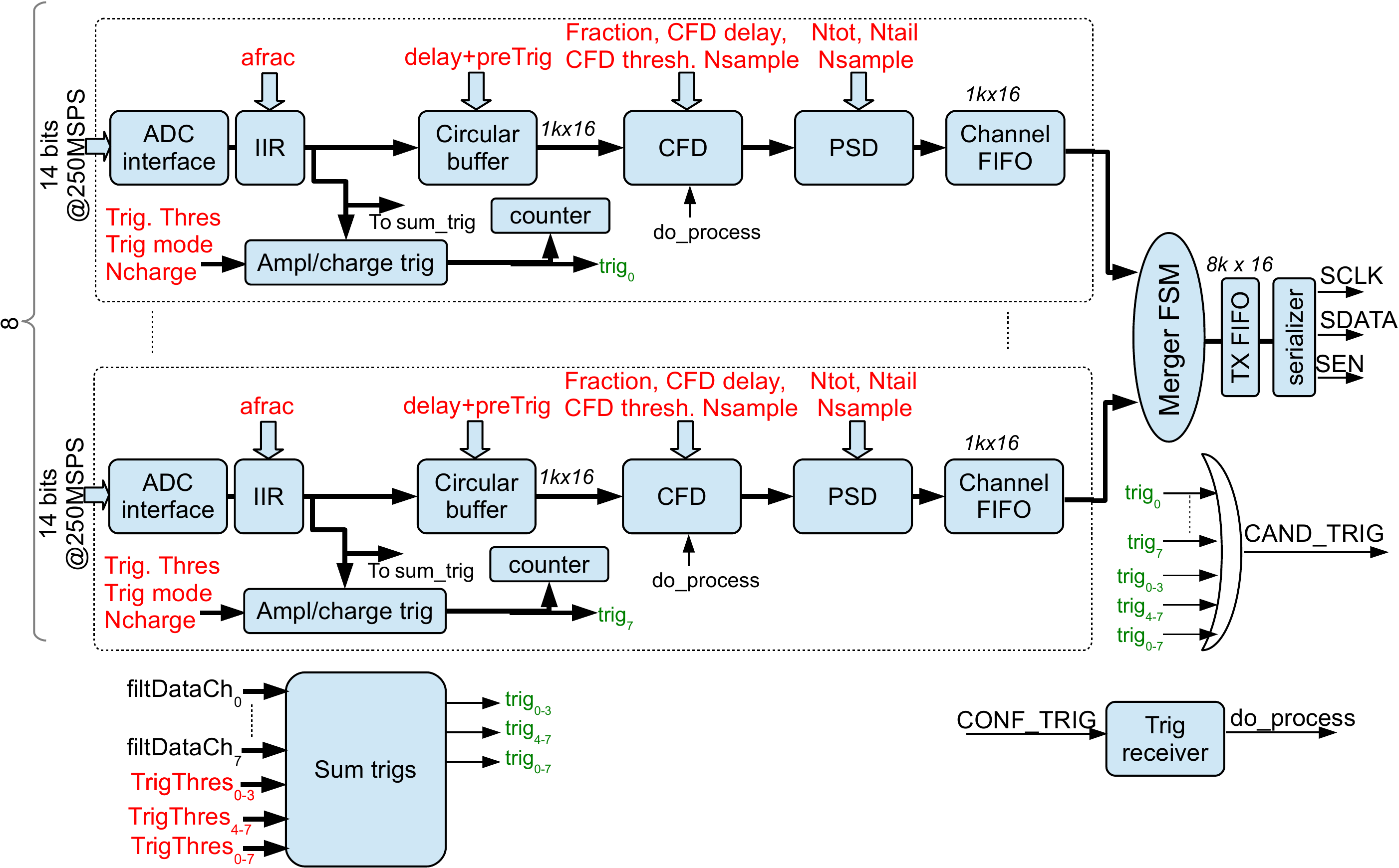}
\caption{Block diagram of the FE8 board firmware. The adjustable parameters are shown in red.}
\label{FE8fw}
\end{center}
\end{figure}

As shown in the FE8 firmware block diagram (fig.~\ref{FE8fw}), the channel processing is done in parallel.
At first, the baseline caused by various offsets (amplifier, ADC) is removed thanks to an Infinite Impulse Response (IIR) first order high pass filter ($\rm f_{-3dB} \simeq 40\,kHz$).
Each filtered signal is sent in parallel to the trigger modules and to a circular buffer.
At the channel level, the trigger can be done on the amplitude or on the Riemann integration over a sliding window of $\rm N_{charge}$ samples ($\rm N_{charge}$ is a configurable parameter).
At the board level, these two kinds of trigger can be generated with the instantaneous sum of 4 or 8 channels.
For each trigger source, a 32-bit counter is implemented for monitoring the trigger rate.

The circular buffer is used to compensate the trigger validation path delay (about 30 clock cycles, i.e. 120\,ns) and to permit pre-triggering.
The data flowing out of the circular buffer are used to fed the CFD that searches the sample numbers ($\rm N_{Zc}$) corresponding to each threshold crossing.
Then the found CFD times ($\rm N_{Zc}$) are used to parametrize the PSD computations according to the sketch shown in fig.~\ref{PSD_1}.
The PSD block provides the computed $\rm Q_{tot}$, $\rm Q_{tail}$, $\rm N_{Zc}$, over-range monitors and, if requested through the ``debug mode'', all the samples used for the computation to the channel FIFO.
Over-range monitors were implemented to flag any miscalculation of $\rm Q_{tot}$ and $\rm Q_{tail}$ due to ADC clipping (over-range).
This is mandatory in ``normal mode'' since the samples used for the charge calculations are no longer available.
All channel FIFO data are aggregated in the TX\_FIFO and transferred serially with a 16-bit custom protocol to the TRB.
The custom synchronous serial protocol, which is operated at 125\,Mbps,  is using port 5/6 for data/enable and TCLKB for clocking.
The trigger channel, operated at 250\,MHz, is hosted by port 7 (candidate and confirmed).

It must be noted that the CFD/PSD processing, which is pipelined, starts only upon the reception of a confirmed trigger.
Therefore, in theory a new confirmed trigger could be accepted every $\rm N_{sample}$ clock cycles (see fig.~\ref{PSD_1}), in practice: 8 additional clock cycles are required in both the normal and debug mode to save the computed data in the channel FIFO and $\rm N_{sample}$ further clock cycles are required in the debug mode.
Consequently, in normal mode with a typical setting $\rm N_{sample}=60$, an instantaneous accepted trigger rate of $\rm \frac{250}{60+8}=3.6\,MHz$ can be reached.

\section{Trigger and readout board}
\begin{figure}
\begin{center}
\includegraphics[angle=0,width=0.49\textwidth]{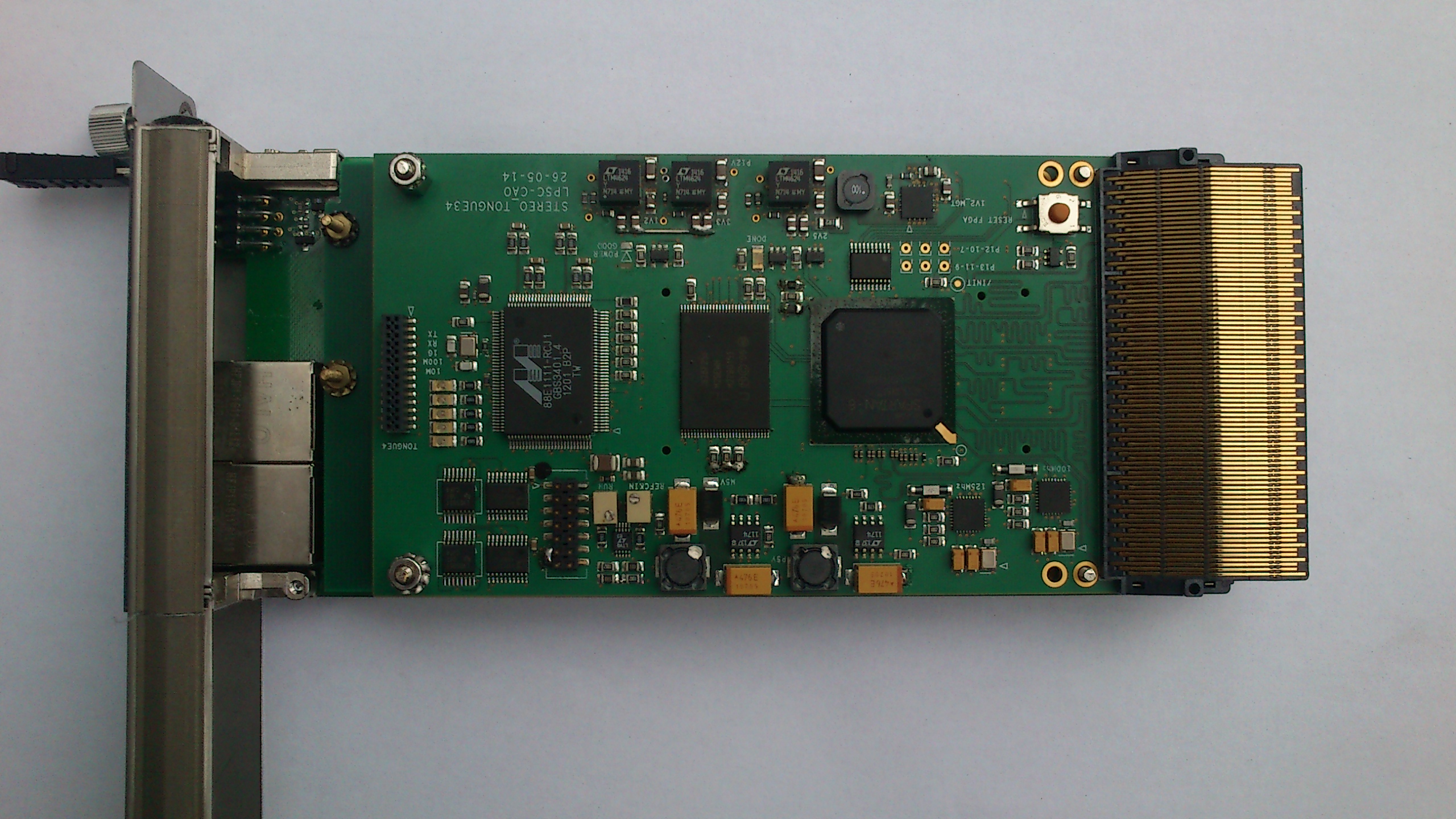}
\includegraphics[angle=0,width=0.49\textwidth]{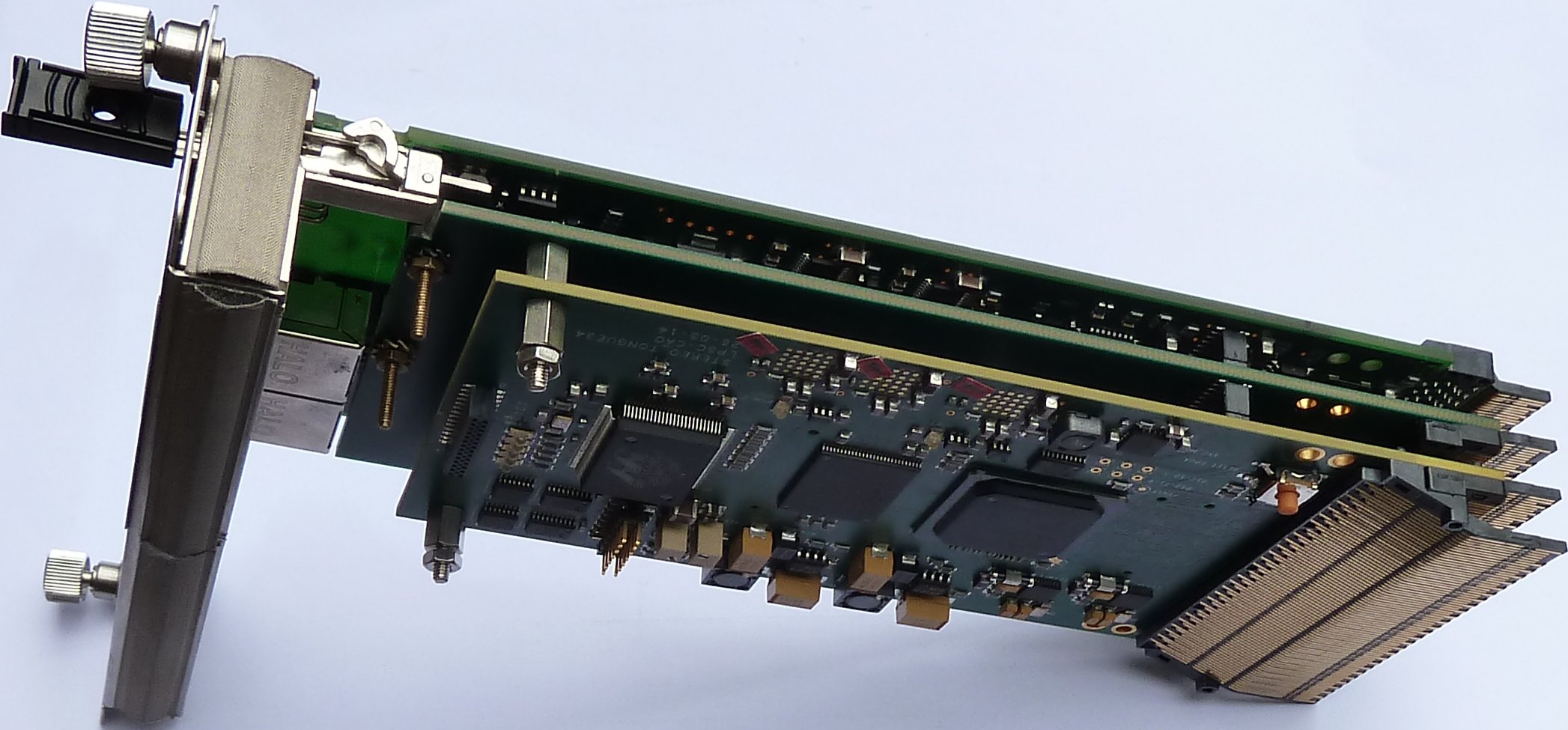}
\caption{Pictures of the two boards composing the TRB mounted on the NAT\textregistered \  MCH. The main TRB board is the top board in the left picture. The extension PCB providing the SMA connectivity with the front panel is not mounted.}
\label{DAQhw}
\end{center}
\end{figure}

In order to benefit from the star architecture of the MTCA, the trigger and readout board was designed as a set of two mezzanine extension boards that are mounted on the NAT\textregistered \  MCH as shown in  fig.~\ref{DAQhw}.
The TRB is connected to the MCH with a connector that provides the power supply and the Ethernet connectivity.
The PCB associated with the second tongue contains only buffers for the clock trees (FCLKA, TCLKA) and the required connectivity for interconnecting the MCH, the tongue 2 and the main TRB board.
The main TRB board is equipped with an FPGA (XC6SLX45T-FGG484) which takes care of the trigger and readout, its associated flash memory for the BPI mode, the power converters and the connectivity with the tongue 3 and 4.
Likewise the FE8, the flash memory can be in-situ updated via IPBus.

\begin{figure}
\begin{center}
\includegraphics[angle=-0,width=0.8\textwidth]{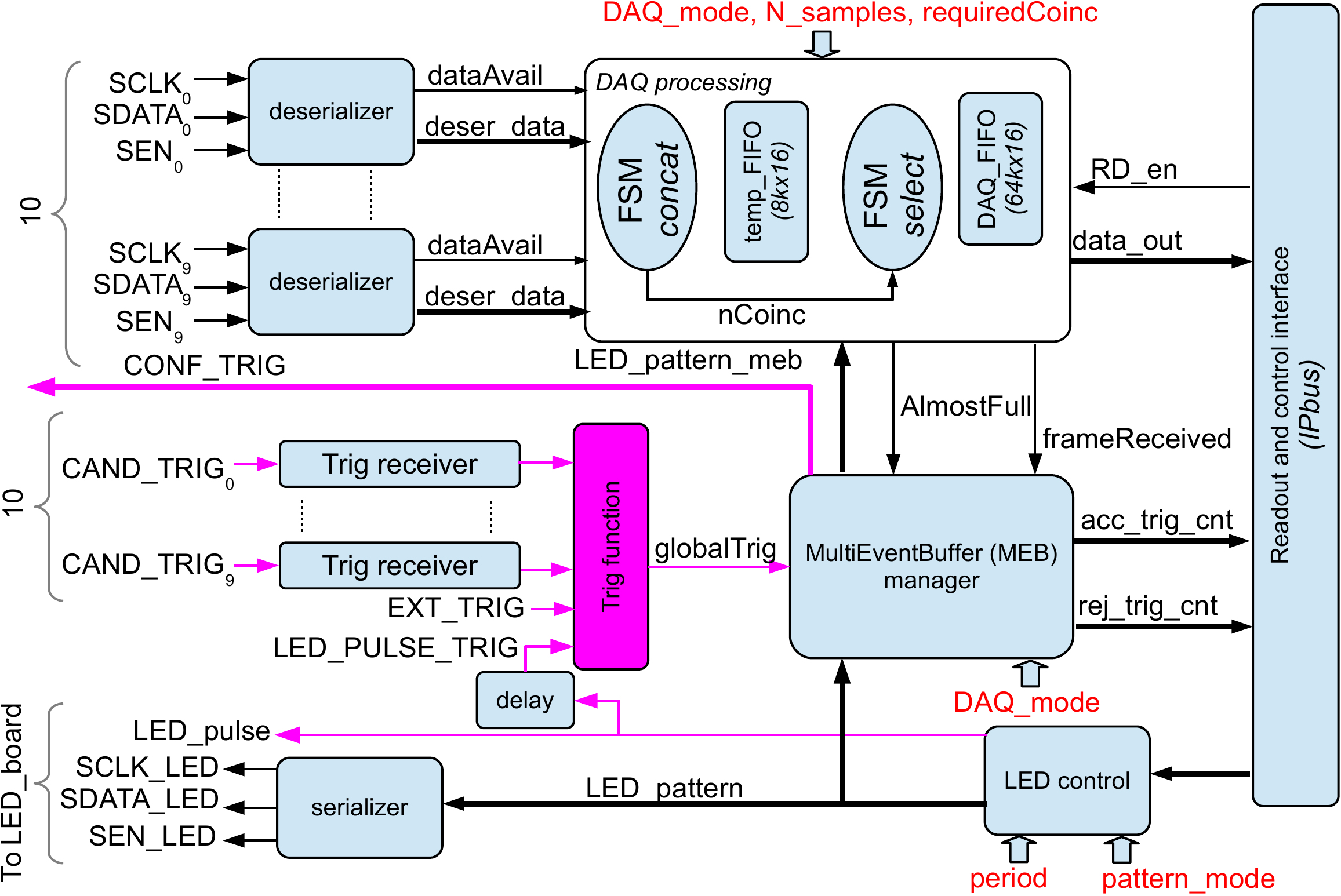}
\caption{Block diagram of the TRB firmware. The three main parts are shown: the LED board controller, the trigger and the readout.}
\label{DAQfw}
\end{center}
\end{figure}

A block diagram detailing the firmware can be seen in fig.~\ref{DAQfw}. 
It is composed of three parts: the LED board controller, the trigger and the readout.

The LED controller is in charge of requesting periodic LED flashes to the LED board according to a preselected pattern (incremental or fixed) of 6 LED. 
The 6-bit pattern is communicated to the LED board by a custom synchronous serial protocol using port 5/6 for data/enable and TCLKB for clocking.
This serial link operates at 125\,Mbps; the 4\,ns LED pulse is sent a few hundredths of nanoseconds after the pattern transmission.
Given the fact that the latencies are fixed and known, a delayed version of the LED pulse is used as a candidate trigger (for T1) to ensure that the good event is recorded by the DAQ.

The trigger part receives the candidate trigger from the ten FE8, the external trigger and a delayed version of the LED pulse. 
These candidate triggers are passed through a trigger function to form the global T1 trigger candidate.
The T1 trigger function is a logical OR of the candidate triggers where a veto condition can be set if candidate triggers are issued from FE8 monitoring the Cerenkov detector.
This latter trigger is accepted by the multi-event buffer only if there is enough space in the readout buffers (TX\_FIFO in FE8 and DAQ\_FIFO in TRB) to store the event, and if the trigger pulses are separated by the minimum time interval, see CFD/PSD processing in section~\ref{FEB} for further details.
Given the condition for accepting the trigger set on the DAQ FIFO ($\rm 64\,k \times 16$) fullness, and the IPBus readout rate, an average accepted trigger rate larger than 1\,kHz can be sustained.
Two 32-bit counters are implemented for monitoring the accepted and rejected candidate T1 triggers.

The readout part is in charge of deserializing the data provided by the ten FE8 boards and to aggregate them.
The aggregated data are then analyzed by the ``\emph{select}'' finite state machine (FSM), that provides a second level (T2) trigger condition, and stored in a temporary buffer.
Eventually, if the T2 condition is met, the event stored in the temporary buffer is transferred to the DAQ\_FIFO and made available for readout.

The T2 trigger definition is still evolving.
Currently a multiplicity condition on the number of hits is implemented.
In the future, total charge conditions on specific parts of the detector could be implemented.
\section{LED board}
\begin{figure}
\begin{center}
\includegraphics[angle=-90,width=0.55\textwidth]{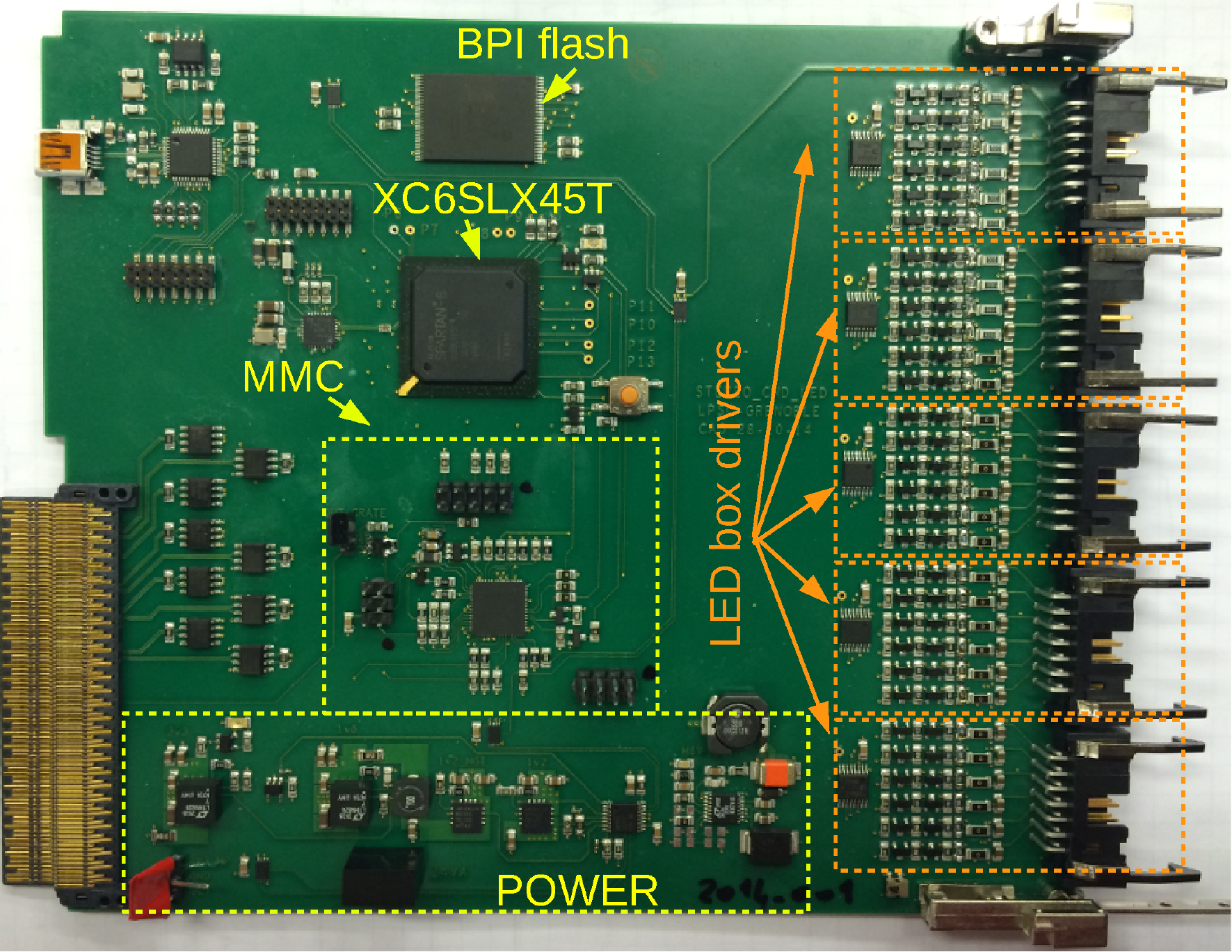}
\caption{Picture of the LED board.\label{LEDboard}}
\end{center}
\end{figure}

\begin{figure}
\begin{center}
\includegraphics[angle=-0,width=0.5\textwidth]{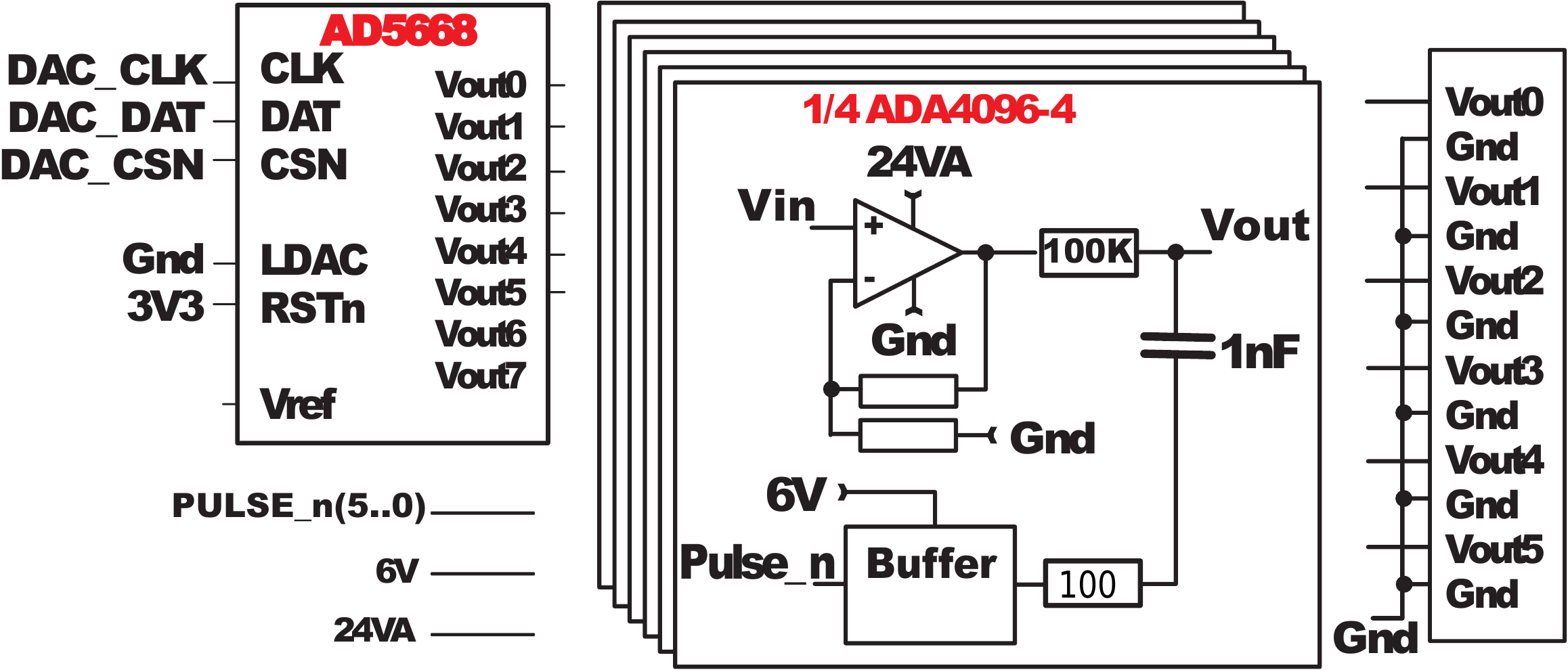}
\includegraphics[angle=-0,width=0.45\textwidth]{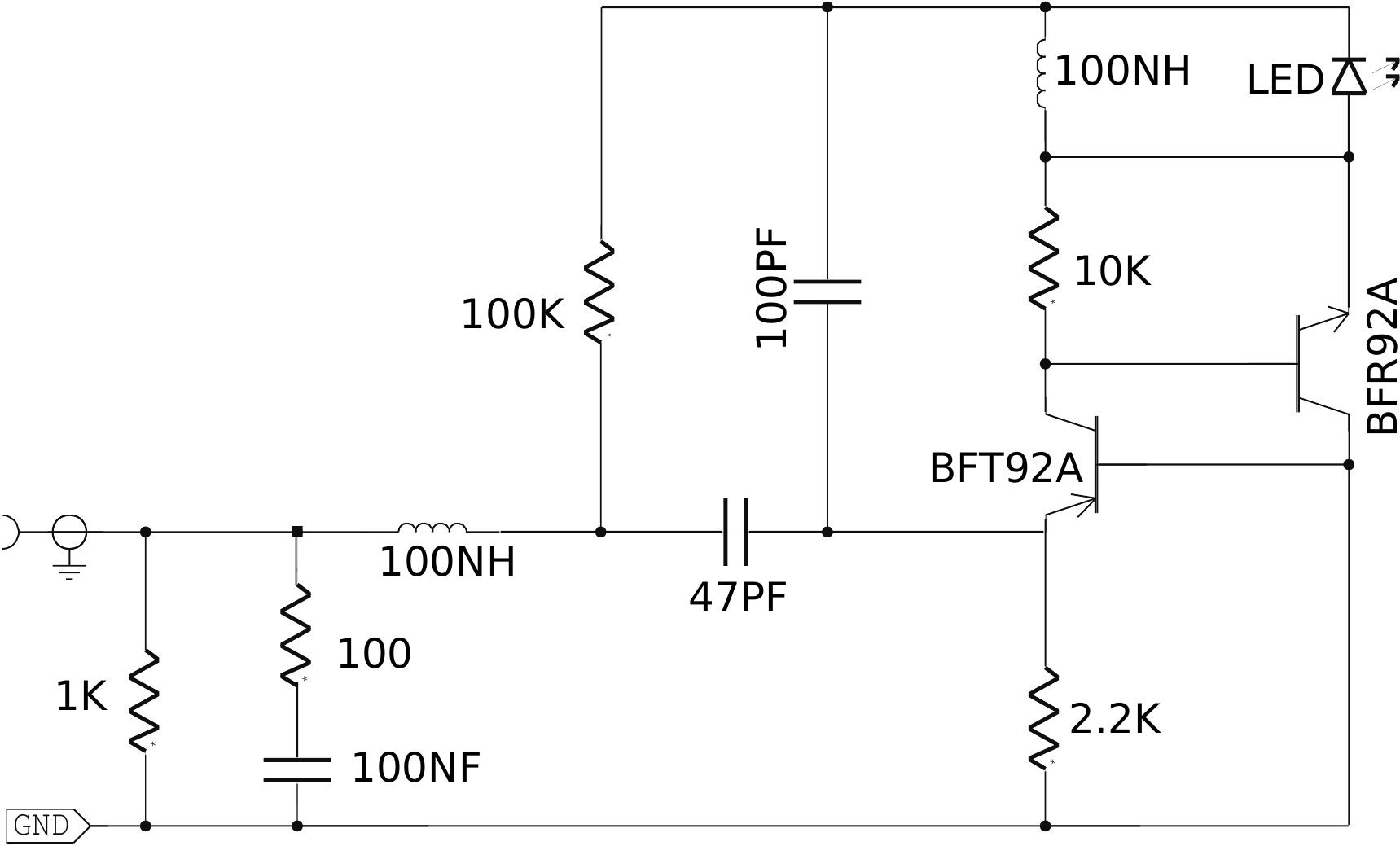}
\caption{Left hand side shows the LED box driver hosted by the LED board. Right hand side is one channel of the LED box.}
\label{LEDhw}
\end{center}
\end{figure}

As introduced in section~\ref{intro}, LED are used to calibrate the detector and to monitor its stability.
The LED system is composed of one LED board (fig.~\ref{LEDboard}) that controls up to five remote LED boxes: three for the inner-detector, one for the Cerenkov detector and one spare. 
Each LED box contains six LED and their associated driving electronics (fig.~\ref{LEDhw}).
The light generated by each LED is injected inside the detector by optical fibers.
To evaluate the detector linearity, the LED can combined by lighting them simultaneously. 
It must be noted, that the LED boxes do not require any power supply.
A single signal pair is used to control each channel: the mean voltage level sets the light level and a superimposed square pulse fires the LED.

The LED board (fig.~\ref{LEDboard}) is built around a FPGA (XC6SLX45T-FGG484) and its associated BPI flash memory, it also features the mandatory MMC module and the five LED box drivers, see fig.~\ref{LEDhw}.
The FPGA permits the slow control of the board, achieved via IPBus, and the on-line control, achieved via the custom serial link, by the TRB.

The slow control part is used to select the LED box driver to use and to adjust the corresponding DAC channels in order to set the mean voltage level (0 to about 22\,V). 
This allows to produce up to 2000 photoelectrons in each PMT.
The on-line part is used to apply in real time and with a known timing the LED pattern by firing the good LED driver channels.
The LED pattern is an incremental pattern covering all possible combinations.

\section{Summary}
In this paper, we have presented the design of the dedicated trigger and acquisition electronics for the STEREO experiment taking place at ILL.
The electronic, which fits in a single microTCA crate, is designed to instrument 68 PMT signals continuously digitized at 250\,MSPS. 
It features two levels of trigger and can record selected data at rate higher than 1\,kHz.

The STEREO electronics is fabricated and was fully tested with cosmic rays using the prototype of the veto-muon detector and in calibration with LED.

\section*{Acknowledgements}
This STEREO collaboration has been funded by the ANR-13-BS05-0007.


\end{document}